\documentclass[]{aa}
\usepackage{graphicx}
\usepackage{times}
\usepackage{natbib}
\bibpunct{(}{)}{;}{a}{}{,} % to follow the A&A style

\begin{document}
  \title{On expansion parallax distances for planetary nebulae}

  \author{Garrelt Mellema\thanks{\emph{Present address:} 
      Netherlands Foundation for Research in Astronomy, P.O. Box 2, 
      NL-7990 AA Dwingeloo, The Netherlands}}

  \institute{
    Sterrewacht Leiden, 
    P.O. Box 9513, 2300 RA, Leiden, 
    The Netherlands\\
    e-mail: mellema@strw.LeidenUniv.nl}

  \offprints{G.~Mellema}

  \date{Received {10 October 2003} / Accepted {4 December 2003}}

  \abstract{The distances to individual wind-driven bubbles such as
    Planetary Nebulae (PNe) can be determined using expansion
    parallaxes: the angular expansion velocity in the sky is compared
    to the radial velocity of gas measured spectroscopically. Since
    the one is a pattern velocity, and the other a matter velocity,
    these are not necessarily the same. Using the jump conditions for
    both shocks and ionization fronts, I show that for typical PNe the
    pattern velocity is 20 to 30 \% larger than the material velocity, and
    the derived distances are therefore typically 20 to 30\% too low. I
    present some corrected distances and suggest approaches to be used
    when deriving distances using expansion parallaxes.

    \keywords{shock waves -- planetary nebulae --- Stars: distances
    --- Stars: AGB and post--AGB --- hydrodynamics}}
  
  \maketitle
%
%________________________________________________________________

\section{Introduction}

Distances to individual nebular objects, such as Planetary Nebulae (PNe) are
important to quantitatively understand their structure and evolution, but
notoriously difficult to determine. Typically uncertainties of a factor of
two can be expected from so-called statistical methods, a review of which can
be found in \citet{1993IAUS..155..109T}.

Probably the best method for measuring the distance to an individual nebula
is the `expansion parallax' method. Here the nebular expansion in the sky as
measured from images taken at different epochs is compared to the radial
velocity as measured spectroscopically. This method became feasible with the
advent of high resolution imaging, in the late 1980's at radio wavelengths
using interferometers \citep{1986ApJ...302L..27M, 1989ApJ...336..294M,
1989ApJ...346..243M, 1993AJ....106.1965H, 1995AJ....109.2600H,
1996ApJ...461..282K, 1996PASP..108..419H, 1998AJ....115.2466C} and nowadays
also in the optical with the Hubble Space Telescope \citep[{\it HST},
][]{1999AJ....118.2430R, 2002AJ....123.2666P, 2002AJ....123.2676L}.

Clearly the simplest application of this method suffers from a number of
drawbacks. One complication is the choice for the spectroscopic velocity,
where different ions often give different values. The other is the assumption
of spherical expansion. However, authors have been making corrections for the
shape and aspherical expansion of the measured nebulae, and applying such
sophisticated templates clearly has made the method more useful, in some
cases reaching claimed errors as low as 10--20\%.

Still, the application requires a reasonably close or particularly rapidly
expanding nebula and to date has only been applied to a limited number of
PNe. Table~1 list all cases with a well determined expansion parallax
distance, leaving out the cases where only upper limits were found.

\begin{table*}
{Table 1 -- Expansion parallaxes}\\
\label{Observations}
\begin{tabular*}{15cm}{*{5}{l}}
\noalign{\smallskip}
\hline\hline
\noalign{\smallskip}
PN & $u_{\rm spectro}$ (km~s$^{-1}$)$^*$ & Distance (kpc) & Radio/Optical & Reference\\
\noalign{\smallskip}
\hline
\noalign{\smallskip}
BD+30~3639 & $22\pm 4$ & $2.8^{+4.7}_{-1.2}$ & r & \citet{1989ApJ...346..243M}\\

BD+30-3639 & $22.0\pm 1.5$ & $2.68\pm 0.81$  & r & \citet{1993AJ....106.1965H}\\

BD+30~3639 & $22\pm 4$ & $1.5\pm 0.4$        & r & \citet{1996ApJ...461..282K}\\

BD+30~3639 & $25.6$ & $1.2\pm 0.12$  & o & \citet{2002AJ....123.2676L}\\

IC~2448    & $17.9\pm 0.3$ & $1.38\pm 0.4$   & o & \citet{2002AJ....123.2666P}\\

%NGC 2392   & $30\pm 10$ & $>1.4$             & r & \citet{1995AJ....109.2600H}\\

NGC 3242   & $26\pm 4$ & $0.42\pm 0.16$      & r & \citet{1995AJ....109.2600H}\\

NGC 6210   & $23\pm 5$ & $1.57\pm 0.40$      & r & \citet{1995AJ....109.2600H}\\

NGC 6543   & $16.4\pm 0.16$ & $1.00\pm 0.27$      & o & \citet{1999AJ....118.2430R} \\

NGC 6572   & $14\pm 4$ & $1.49\pm 0.62$      & r & \citet{1995AJ....109.2600H}\\

NGC 6572   & $14\pm 4$ & $1.2\pm 0.4$        & r & \citet{1996ApJ...461..282K} \\

NGC~6578   & $19.2\pm 0.5$ & $2.00\pm 0.5$   & o & \citet{2002AJ....123.2666P} \\

NGC~6884   & $16.6\pm 0.4$ & $2.20\pm 0.8$   & o &\citet{2002AJ....123.2666P} \\ 

NGC 7027   & $21$          & $0.94\pm 0.2$   & r & \citet{1986ApJ...302L..27M}\\

NGC 7027   & $17.5\pm 1.5$ & $0.88\pm 0.15$  & r & \citet{1989ApJ...336..294M}\\

NGC 7027   & $17.5\pm 1.5$ & $0.703\pm 0.095$ & r & \citet{1993AJ....106.1965H}\\

NGC 7662   & $21\pm 7$     & $0.79\pm 0.75$   & r & \citet{1996PASP..108..419H}\\

VY 2-2     & $19.5\pm 0.4$ & $3.6\pm 0.4$     & r & \citet{1998AJ....115.2466C}\\
\noalign{\smallskip}
\hline
\noalign{\small $^*$ This is the spectroscopic velocity used by the authors, which can
correspond to a shock or ionization front velocity}
\end{tabular*}
\end{table*}

Among the corrections needed to accurately use the expansion parallax method
there is one which to date has received little attention in the literature,
even though it is quite basic. This is the fact that the expansion velocity
as measured in the sky is a pattern velocity, whereas the spectroscopically
measured velocity is a material velocity, and the two are generally not the
same. This effect was touched upon by \citet{1993IAUS..155..315M} and 
\citet{1997IAUS..180..368S}, but has not been taken into account in any
of the published expansion parallax distances.

The extreme case would be that of an R-type ionization front making its way
through a stationary medium, and clearly here the expansion parallax method
becomes useless, as has been mentioned by various authors. Luckily, this
situation is thought to be rare in PNe. However, also in the more common
cases of shock fronts or slower moving D-type ionization fronts, the two
velocities will differ. Although typically not by much, the effect is
systematic, not random, and it should therefore be taken into account when
using the expansion parallax method. Especially as the measurements are 
becoming more accurate over time, this factor can no longer be neglected.

In this paper I calculate the magnitude of the discrepancy between the two
velocities for shocks (Sect.~2), ionization front structures (Sect.~3) and
so-called `shells' (Sect.~4). In Sect.~5 I apply the derived corrections to
the published results of the expansion parallax method and suggest strategies
to optimize the expansion parallax method.

\begin{figure}
\centerline{\includegraphics[width=8cm]{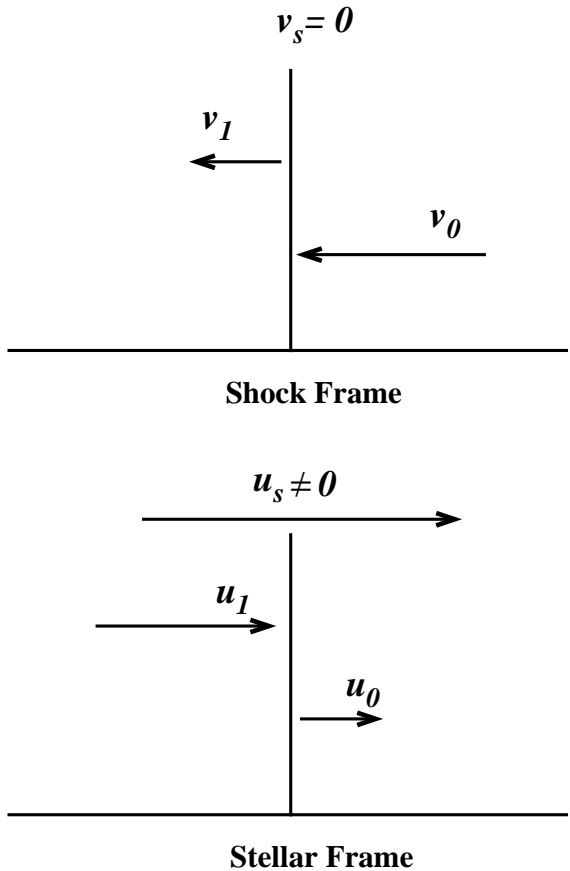}}
\caption{Sketch of the shock configuration and the definition of the
various velocities.}
\label{shocksketch}
\end{figure}

\section{Shock waves}

Since PNe are basically wind-driven bubbles, they contain shocks. Numerical
hydrodynamic modelling has shown that in round or elliptical nebulae the
inner bright rim is normally associated with the shock wave being driven into
the surrounding material by the stellar wind. Key papers describing the
radiation-hydrodynamic evolution of PNe are \citet{1991A&A...248..590M,
1994A&A...290..915M, 1995MNRAS.277..173M}, and a review is presented in
\citet{SchonStef}. The application to an individual PN is shown in
\citet{2000A&A...354.1071C}. Whenever I mention numerical simulations in what
follows, I refer to these papers.

%For NGC 6543 Chandra imaging
%shows that ellipse E25 \citep[see Fig.~1 of] [] {1999AJ....118.2430R}, is the
%wind-driven bubble \citep{2001ApJ...553L..69C}.

The jumps of density and velocity across a shock are given by the
Rankine-Hugoniot conditions. These are usually given for the reference frame
of the shock, in which case the velocity jump is
\begin{equation}
  {v_0 \over v_1} = {(\gamma +1){\cal M}^2 \over (\gamma -1){\cal M}^2 +
  2}\,,
  \label{vjumpshock}
\end{equation}
which for infinite Mach number ${\cal M}$ and an adiabatic index $\gamma=5/3$
gives the classical value of 4. The Mach number of the shock is given by
\begin{equation}
  {\cal M}={|v_0| \over a_0}\,,
  \label{mach}
\end{equation}
where $a_0$ is the sound speed in the pre-shock gas, { given by
\begin{equation}
  a_0=\sqrt{{\gamma kT_0 \over \mu m_{\rm H}}}\,,
  \label{soundspeed}
\end{equation}
with $k$ the Boltzmann constant, $T_0$ the gas temperature, $\mu$ the
mean molecular weight, and $m_{\rm H}$ the mass of atomic hydrogen.}

However, in the stellar frame, the shock has a velocity which we will
call $u_{\rm s}$ and the pre- and post-shock velocities are given by
$u_{0,1}=v_{0,1}+u_{\rm s}$, see Fig.~\ref{shocksketch}.

The expansion parallax method measures $u_s$ from the angular expansion of
the shock front, $\dot\theta = u_s/D$, and $u_1$ from the spectroscopy, and
derives the distance $D$ from the ratio of the two, assuming $u_s=u_1$. From
the expression above it is immediately clear that $u_s$ and $u_1$ are not
equal. If the PN woud be expanding into complete vacuum the two velocities
would be identical, but in the case where it is moving into a surrounding
medium, it is more proper to view the expansion as a disturbance or wave
travelling through the medium, raising its density and velocity. The speed at
which the wave travels is not the same as the velocity to which the material
is accelerated by the passage of the wave.

Let us consider the ratio ${\cal R}$ of the two velocities, $u_{\rm s}/u_1$.
The distance to the PN can then be expressed as ${\cal R}u_1/\dot\theta$.
Using the relations between $u_{0,1}$ and $v_{0,1}$ together with
Eqs.~\ref{vjumpshock} and \ref{mach} (realizing that in our shock frame both
$v_0$ and $v_1$ are negative) gives
\begin{equation}
  {\cal R}={ 
    (\gamma +1 ){\cal M}u_0 + (\gamma + 1){\cal M}^2a_0 \over
    (\gamma + 1 ){\cal M}u_0 + 2({\cal M}^2-1)a_0
  }\,.
  \label{rshock}
\end{equation}
The limits of this ratio are $(u_0+a_0)/u_0$ for ${\cal M} \rightarrow 1$ and
$(\gamma +1)/2$ for ${\cal M} \rightarrow \infty$. This shows that only for
isothermal ($\gamma=1$) hypersonic shocks the ratio tends to one { for
high Mach numbers. Choosing values for the pre-shock velocity $u_0$, pre-shock
sound speed $a_0$, and shock Mach number ${\cal M}$ gives a value for ${\cal
R}$. The material velocity $u_1$ can then be found from
\begin{equation}
  u_1=u_0+2a_0\left({{\cal M}^2-1 \over (\gamma+1){\cal M}}\right)\,.
  \label{u1}
\end{equation}
Reversely, given values for $u_0$, $u_1$, and $a_0$, the shock's Mach number
can be found from
\begin{equation}
  {\cal M}={(\gamma +1)(u_1-u_0)+\sqrt{(\gamma+1)^2(u_0-u_1)^2+16a_0^2}
      \over 4a_0}\,,
  \label{machfromvelocities}
\end{equation}
which can then be used to derive ${\cal R}$ using Eq.~\ref{rshock}.}

\begin{figure}
\includegraphics[height=62mm,clip=]{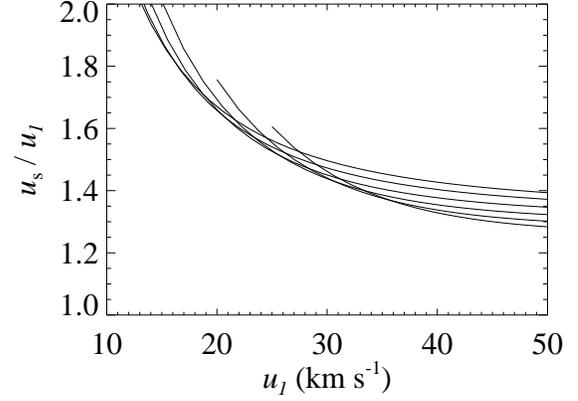}
\caption{The ratio ${\cal R}$ ($=u_{\rm s}/u_1$) as function of $u_1$ for the
case $\gamma=5/3$ (no cooling), and (adiabatic) sound speed
$a_0=15$~km~s$^{-1}$. The six curves correspond to different values for the
pre-shock velocity $u_0$: 1 (highest curve at $u_1=50$~km~s$^{-1}$), 5, 10,
15, 20, and 25 (lowest curve) km~s$^{-1}$.}
\label{shockratios-adiab}
\end{figure}

\begin{figure}
\includegraphics[height=62mm,clip=]{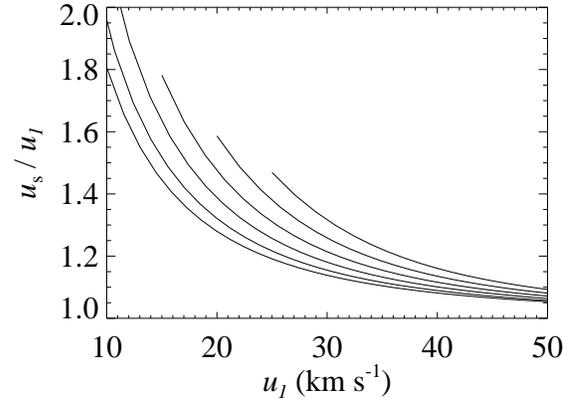}
\caption{The ratio ${\cal R}$ ($=u_{\rm s}/u_1$) as function of $u_1$ for the
isothermal case ($\gamma=1$), and (isothermal) sound speed
$a_0=11.7$~km~s$^{-1}$. The six curves correspond to different values for the
pre-shock velocity $u_0$: 1 (lowest curve), 5, 10, 15, 20, and 25 (highest
curve) km~s$^{-1}$.}
\label{shockratios-isothermal}
\end{figure}

In Fig.~\ref{shockratios-adiab} I plot the ratio ${\cal R}$ as function of
$u_1$, for six pre-shock velocities $u_0$ in the range 1 to 25~km~s$^{-1}$,
using a $\gamma$ value of 5/3 (monatomic ideal gas), { and
$a_0=15$~km~s$^{-1}$, the adiabatic sound speed for an ionized gas at
$10^4$~K}. Most PNe have high
densities and relatively slow shocks, so that the shocks are expected to be
isothermal. Using $\gamma=1$ is therefore appropriate, and in
Fig.~\ref{shockratios-isothermal} I show ${\cal R}$ as function of $u_1$ for
this isothermal case, { where I have used an isothermal sound speed of
11.7~km~s$^{-1}$, valid for an ionized gas of cosmic abundances at a
temperature of $10^4$~K. As expected from the limits derived above,} the
ratio is largest for low velocities. { I should note that the ratio
does depend on the choice for the sound speed $a_0$. A value corresponding
to an electron temperature of $10^4$~K is typical for PNe, but the
temperature can be both higher and lower than this, ranging from 5000
to 15,000~K. Lower temperatures give lower values of ${\cal R}$.}

Looking at the data in Table~1, it is clear that for these PNe, we are never
in the very high Mach number regime. Using Fig.~\ref{shockratios-isothermal}
one can see that for the observed ranges of velocities and for $\gamma=1$,
the typical ratio ${\cal R}$ is between 1.3 and 1.5, although it can be as
high as 1.8. Obviously this falls outside the formal errors of the
method and becomes as important an effect as the geometric corrections
applied for example by \citet{2002AJ....123.2676L}. Furthermore, the factor
is {\it always}\/ larger than 1, so it does not make sense to add a 20-30\%
extra uncertainty to the distances. Rather, the distances should be scaled up
by 30\% and an extra error of $\sim 10$\% added to it.
Figure~\ref{shockratios-isothermal} can be used to estimate the magnitude of
this effect for individual PNe, which I will do in Sect.~5.

The conclusion then is that if the expansion parallaxes are measured
from the shocked component of the PN, the distances should be
multiplied by a factor ${\cal R}$ given by Eq.~\ref{rshock}, where
uncertainties in the velocity of the material into which the shock is
expanding adds a $\sim$10\% error. I will come back to this point in
Sect.~5.

\section{Ionization fronts}

Ionization fronts are either R- or D-type. The basic difference is the speed
of the front which for R-type fronts is higher than twice the sound speed in
the neutral gas, giving the gas no time to react to the presence of the
ionization front, wheras for D-type it is lower than twice the sound speed in
the neutral gas. The result is that D-type fronts actually consist of a
combination of an ionization front and a preceding shock front. A good
overview of basic ionization front theory can be found in
\citet{1992phas.book.....S}.

The numerical simulations for the formation of PNe which studied the effects
of ionization fronts in detail, showed that although R-type fronts occur, they
never persist for long, and it would be highly unlikely to observe one in
action. When they occur, they are of the so-called weak type, {i.e.\ }
supersonic with respect to both the neutral and ionized flow regions.

Weak D-type fronts { (moving subsonically with respect to both the ionized
and neutral regions)} are more commonly found in the models, and persist for
a longer time, and could be present in a number of observed PNe, although it
is generally hard to prove this. The simulations also show that the presence
of a D-type front sets up a disturbance in the gas which persists long after
the front itself has disappeared, creating the shells which I discuss in the
next section.

Basic ionization front theory shows that the velocities $v_1$ and $v_2$ on
either side of the front are related by
\begin{equation}
  { v_1 \over v_2 } = {1 \over 2a_2^2} 
  \left( a_1^2+v_1^2 \pm 
    \left[ (a_1^2+v_1^2)^2 - 4a_2^2v_1^2 \right]^{1/2} \right)\,.
  \label{vjumpion}
\end{equation}
The indices 1 and 2 refer to the neutral and ionized sides, respectively.
Just as in the previous section, $v$ refers to velocities in the frame of the
discontinuity, and $u$ to velocities in the stellar frame. If the front moves
with a velocity $u_{\rm i}$ in the stellar frame, the velocities on either
side of it are $u_{1,2}=v_{1,2}+u_{\rm i}$. For the weak R-type fronts one
has to use the $-$ sign in Eq.~\ref{vjumpion} and for the weak D-type fronts
the $+$. The velocity $v_1$ has to be larger than $v_{\rm R} =
a_2+(a_2^2-a_1^2)^{1/2}$ for R-type fronts and smaller than $v_{\rm D}=
a_2-(a_2^2-a_1^2)^{1/2}$ for D-type fronts.

\begin{figure}
\centerline{\includegraphics[height=62mm,clip=]{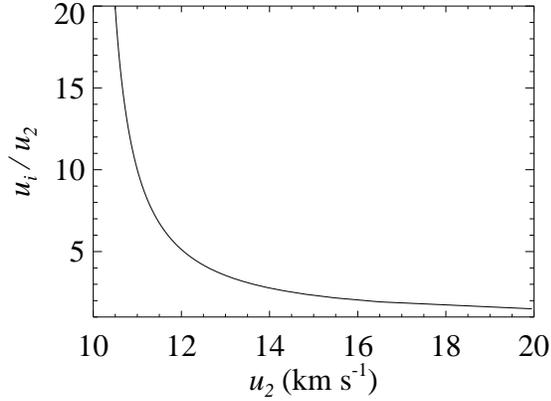}}
\caption{The ratio ${\cal R}$ ($=u_{\rm i}/u_2$) as function of $u_2$ for
a weak R-type ionization front running into a 10 km~s$^{-1}$ wind. The
weaker the front the closer the ratio gets to one.}
\label{ifront_rw}
\end{figure}

In Fig.~\ref{ifront_rw} I show the ratio ${\cal R}$, now defined as $u_{\rm
i}/u_2$, against $u_2$ for weak R-fronts running into a medium with $u_1 =
10$~km~s$^{-1}$. { For the sound speeds I chose $a_1=1$~km~s$^{-1}$ and
$a_2=10$~km~s$^{-1}$, and tests show that the ratio ${\cal R}$ is hardly
sensitive to this choice.} As can be expected this ratio is large
for very fast moving fronts, and approaches 1 for the slower moving
ones. Because of the large range of ratios, it would seem dangerous to apply
any type of correction in case of R-type ionization fronts.

\begin{figure}
\centerline{\includegraphics[height=62mm,clip=]{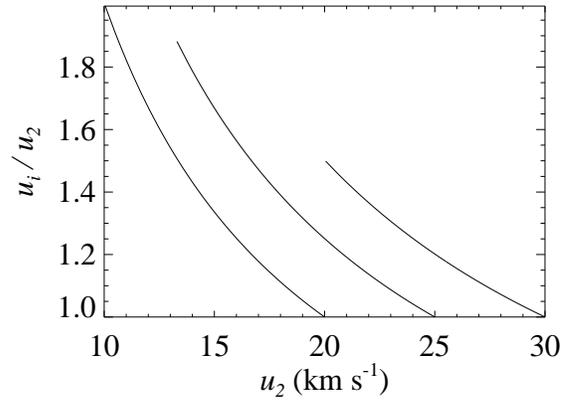}}
\caption{The ratio ${\cal R}$ ($=u_{\rm i}/u_2$) as function of $u_2$ for
a weak D-type ionization front running into a wind with velocity $u_1$ of
(from left to right) 20, 25, and 30 km~s$^{-1}$.}
\label{ifront_dw}
\end{figure}

In Fig.~\ref{ifront_dw} I show the same ratio ${\cal R}$ against $u_2$ for
weak D-fronts, a situation which is more likely to occur in real PNe. Normally
the shock front which precedes the D-front will have accelerated the AGB wind
to a higher velocity. The numerical simulations show this velocity to be
typically around $u_1=20$~km~s$^{-1}$. I plot the ratio for three
values of $u_1$. Interestingly, around the typically observed velocities, the
correction is of the same order of magnitude as for the shock waves in
Sect.~2.

The full solution for a D-type front would eliminate the choice for
$u_1$.  \citet{1992phas.book.....S} showed how this can be done for a
constant flux of ionizing photons impinging on a static and constant
density environment. Attempts to do this for a stellar wind type
environment have been only partly successful. Giuliani (1989) found a
self-similar solution which requires the flux of ionizing photons to
be time dependent and fall off as $t^{-1}$, which does not apply to
most PNe.

\citet{1986ApJ...302L..27M, 1989ApJ...336..294M, 1989ApJ...346..243M}
applied a series of corrections to his determination of the distance
to NGC 7027, BD+30~3639 and NGC 6591, which are partly related to the
difference between the material velocity (which he referred to as the
``bulk velocity''), and the (pattern) speed of the ionization front.
The magnitudes of these corrections are actually similar to the ones
found above, 1.1 to 1.2. Perhaps somewhat confusingly he applies these
corrections to the measured angular expansion rate, so that the final
figures he quotes for $\dot\theta$ are not actually the ones
measured. He derives this correction from the fact that as the ionized
shell expands, the density will go down, reducing the number of
photons used up in recombinations, and allowing the ionization front
to expand. This is a different approach to the one used above and does
not use the jump conditions. However, the basic idea is still that the
ionization front expands faster than the material flow of the gas, so
the corrections are related. In not considering the jump conditions he
implicitly assumes a weak R-type ionization front with a small density
jump, and the correction corresponds therefore to the one from
Fig.~\ref{ifront_rw}. The numerical simulations as well as the
observed velocities show that this is not correct, although the
correction factors come out at similar values. Note that Masson is the
only author to actually apply {\it any}\/ type of correction. In
Sect.~5 I will comment some more on this.

The conclusion from this section is that it is more difficult to find a {
general} solution for ionization fronts, which makes them less suitable than
shock fronts for applying the expansion parallax method. However, the results
in this section allow for corrections to be made, and below I will illustrate
how these corrections could be applied.

\section{Shells}

Numerical simulations show that the density disturbance induced by the D-type
ionization front will persist and can be identified with the observed
attached shells around the bright core nebula, such as for example in
NGC~3242.  \citet{2003MNRAS.340..417C} compiled a list of PNe with haloes and
shells, which can serve as a reference. We follow their nomenclature and
refer to these structures as {\it shells}.

Analytical models for the expansion of such shells are available in
\citet{1997ApJ...488..263C} and \citet{2002ApJ...580..969S}, even though the
latter had a different application in mind. Using their results, it is
possible to extract a similar ratio ${\cal R}$ of pattern speed over material
speed, where the pattern speed is now the movement of the edge of the shell.
Since these models actually use the isothermal shock conditions, the results
are basically identical to ones from Sect.~2. However, I list them
separately, since shells are commonly observed and the
\citet{1997ApJ...488..263C} and \citet{2002ApJ...580..969S} results are the
full solutions including the acceleration of the AGB wind after ionization.

\begin{figure}
\centerline{\includegraphics[height=62mm,clip=]{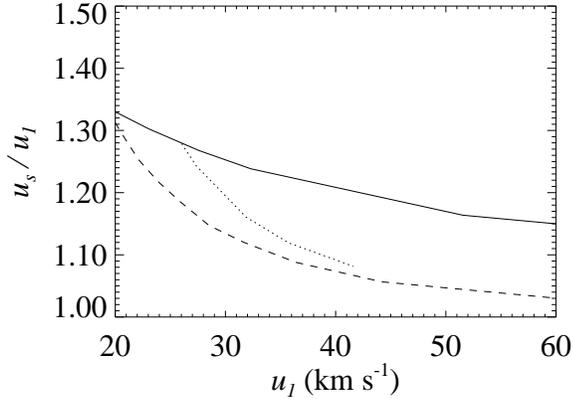}}
\caption{The ratio ${\cal R}$ ($=u_{\rm s}/u_1$) as function of $u_1$
for the shell solutions from \citet{1997ApJ...488..263C} and
\citet{2002ApJ...580..969S}. The solid line shows the solutions for a
photo-ionized wind, the dotted line the same but with the influence of
a wind included. The dashed line shows the result for photo-ionized
winds with different density laws steeper than $r^{-2}$.}
\label{shells}
\end{figure}

For an AGB wind of constant mass loss and velocity, the results from
\citet{1997ApJ...488..263C} show that ${\cal R}$ is between 1.10 and 1.34. I
plot this ratio against $u_1$ in Fig.~\ref{shells}, together with the ratio
from his solution including the effect of a stellar wind, which gives
somewhat lower values. The ratio for an AGB wind density falling off steeper
\citep[$\rho \propto r^{-\alpha}$ with $\alpha>2$,][]{2002ApJ...580..969S}
are also shown.

The numerical simulations also show that when an attached shell forms, its
velocity structure is that of a rarefaction wave, with a positive outward
gradient. This means that the bright rim expanding into this will be moving
into an area with a velocity lower than the original AGB wind velocity. The
implication for the expansion parallax method is that when using the rims of
PNe with attached shells a value of $u_0$ lower than 10~km~s$^{-1}$ is more
appropriate when determining the correction factor from Sect.~2. 

The conclusion from this section is that for attached shells a correction of
around 20\% is needed.

\section{Implications for distances and the method}

The results presented thus far show that for the published distances, the
results should typically be scaled up with a factor $1.3 \pm 0.2$, which
interestingly enough eliminates the systematic discrepancy between the
expansion parallax distances and statistical distances noted by
\citet{2002AJ....123.2666P}. In this section I will go through the list of
PNe for which distances have been measured, consider corrections for each
case, and give some suggestions on how to improve the usage of the expansion
parallax method further.  { For calculating the corrections I use the
values $\gamma=1$ and $a_0=11.7$~km~s$^{-1}$ for the cases of shocks
(corresponding to Fig.~\ref{shockratios-isothermal}), $a_1=1$ and
$a_2=10$~km~s$^{-1}$ for ionization fronts (corresponding to
Fig.~\ref{ifront_dw}). The first part of this section is intended not only to
derive new distances, but also to illustrate how to derive and use the
correction factors.}

\begin{table}
{Table 2 -- New distances}\\
\label{newdistances}
\begin{tabular*}{8cm}{*{2}{l}}
\noalign{\smallskip}
\hline\hline
\noalign{\smallskip}
PN & Distance (kpc) \\
\noalign{\smallskip}
\hline
\noalign{\smallskip}
BD+30~3639 & $1.3\pm 0.2$\\

IC~2448    & $2.07\pm 0.62$\\

NGC 3242   & $0.55\pm 0.23$\\

NGC 6543   & $1.55\pm 0.44$\\

NGC~6578   & $2.90\pm 0.78$\\

NGC~6884   & $3.30\pm 1.24$\\

NGC 7027   & $0.68\pm 0.17$\\

NGC 7662   & $1.19\pm 1.15$\\

VY 2-2     & $4.68\pm 1.20$\\
\noalign{\smallskip}
\hline
\end{tabular*}
\end{table}

\subsection{New distance estimates}

Going through the list there are a number of PNe which contain bright rims
and attached shells, and the rim can therefore be considered to be bounded by
a shock. These are NGC 3242, NGC 6578, NGC 6884, NGC 7662, and IC 2448. {
Making conservative assumptions about the value of $u_0$ ($13\pm
12$~km~s$^{-1}$) and using the results from Sect.~2, the distances to these
PNe should be increased by factors $1.3\pm 0.2$, $1.45\pm 0.15$, $1.5 \pm
0.15$, $1.5\pm 0.3$, and $1.5\pm 0.1$ respectively. The uncertainties are
{ due partly to the uncertainties in the reported spectroscopic velocities
and partly} to the uncertainty in the value of $u_0$. The latter could in
principle be reduced if one assumes that the velocities in the shells follow
the numerical models, in which case $u_0$ will be low ($<10$~km~s$^{-1}$).}
I list the new (conservative) values in Table~2.

NGC 6543 is a more complex nebula. The Chandra results show that it
contains a wind-driven bubble, but the ionized material surrounding this
bubble does not resemble a standard shell. This complicates the choice for
$u_0$, but within reasonable limits the correction factor is $1.55\pm 0.15$.

Three PNe in Table~1 are candidates for the presence of an ionization front:
BD+30~3639, NGC 7027, and Vy2-2. 

Of these BD+30~3639 is a single shell, and relatively young PN with molecular
emission around it, although the deeper exposures also seem to show an
ionized component around the bright shell \citep{1998AJ....116.1357S}.
\citet{1996ApJ...461..282K} used a detailed analysis of the expansion of this
object in order to derive the distance, assuming we are observing an
ionization front. Part of this analysis is a correction factor of 1.19 to
account for the difference between the material and pattern velocities, as
mentioned in Sect.~3. Using the results from Sect.~3 for a weak D-type
front, I estimate a correction factor of $1.2\pm 0.2$ assuming a velocity of
$25\pm 5$~km~s$^{-1}$ for the shocked neutral gas and a spectroscopic
velocity of $22\pm 4$~km~s$^{-1}$ \citep[as used by][] {1996ApJ...461..282K},
confirming their result for the distance.

Li et al. used optical {\it HST}\/ data, better spectroscopic data and a
correction for the ellipticity of the nebula to arrive at a smaller
distance. Using their preferred value for the spectroscopic velocity,
25.6~km~s$^{-1}$, I find a correction factor of $1.1\pm 0.1$ if we are seeing
an ionization front running into a shocked component with a velocity of 25.6
-- 30~km~s$^{-1}$, and a correction of $1.3\pm 0.1$ if we are seeing a shock.
It therefore seems reasonable to increase the distance by 10\% and to double
the uncertainty on that number: $1.3\pm 0.2$~kpc, bringing it closer to the
value from \citet{1996ApJ...461..282K}

%Since his type of analysis includes a model of a shell, it is hard
%to compare to the simpler discontinuity approach used here. I would therefore
%not apply any further corrections to the derived distances.

NGC 7027 is a more complex PN where an ionization front has to be present,
given the large amounts of dust and molecules surrounding this object. The
elliptical shell seen at radio wavelengths (and partly obscured at optical
wavelengths) could be the ionization front and is analyzed as such by
\citet{1986ApJ...302L..27M,1989ApJ...336..294M}. To correct for the
discrepancy between spectroscopic and angular velocities, he uses a factor of
1.2 in the first and 1.15 in the second paper (note that the second paper
quotes 1/1.15=0.83 as the correction factor). Further corrections for the
decrease of the radio flux in the second paper actually largely cancel the
effect of this factor.

\citet{2003MNRAS.340..381B} analyzed high resolution optical long slit
spectra of NGC 7027 and from the [O~III] line derived an equatorial expansion
velocity considerably lower than used by \citet{1989ApJ...336..294M}, namely
$13 \pm 1$~km~s$^{-1}$. \citet{2002A&A...384..603C} report K-band imaging and
spectroscopy for NGC~7027. Their best fitting model has an equatorial
expansion velocity in Br$\gamma$ of about 13~km~s$^{-1}$ (Cox \& Huggins,
private communication; this number is not given in the paper), consistent
with the [O~III] value. For the molecular emission (H$_2$) the same authors
find an equatorial velocity of $\sim$15~km~s$^{-1}$. Allowing all possible
values for the velocity of the neutral material $u_1$ (13 --
23~km~s$^{-1}$), the correction factor would be $1.4\pm 0.4$. Assuming that
the H$_2$ expansion is indicative of the value of $u_1$ (13 --
17~km~s$^{-1}$), the ratio becomes $1.2\pm 0.2$. If the ionized shell is
actually bounded by a shock, the results of Sect.~2 show that the correction
factor would be $1.75\pm 0.15$ for all allowed values of the pre-shock velocity
$u_0$.

Taking an angular expansion of $4.84\pm 0.82$ mas~yr$^{-1}$ \citep[][this
includes the correction for flux variations $0.47\pm 0.47$), but not the
correction factor of 0.83]{1989ApJ...336..294M}, combining this with an
equatorial velocity of 13~km~s$^{-1}$, and applying a correction of $1.2\pm
0.2$, I arrive at a new distance of $680\pm 170$~pc. Note that this error is
rather optimistic in view of the wide range of correction factors mentioned
above.

Vy2-2 is a compact PN which is hard to categorize. Also here one may have to
be aware of ionization fronts since this is a fairly young, low excitation
PN. Assuming either an ionization front or an isothermal shock, the reported
distance should be scaled up by $1.3\pm 0.3$

Two PNe in the list are hard to categorize in the standard wind-blown bubble
plus photo-ionization scheme: NGC 6210 and NGC 6572. I therefore will not
suggest any corrections for these, although shocks could very well be
present. For NGC 6210 this would imply a 30\% increase in the reported
distance, but for NGC 6572 the correction could become very large (1.4--1.8)
due to the low value of the reported spectroscopic velocity.

\subsection{Improving the method}

{ The correction factors derived in this paper depend on a number of
parameters. As already indicated, due to the isothermal nature of the slow
shocks in PNe, the preferred value for the adiabatic index $\gamma$ is 1.
The sound speed in the ionized medium ($a_0$) I have taken to be
11.7~km~s$^{-1}$ in Fig.~\ref{shockratios-isothermal}, which is the value for
an electron temperature of $10^4$~K. If the PN is known to have a
particularly high or low electron temperature the figure should be
recalculated for the appropriate sound speed, since the values for ${\cal R}$
unfortunately do depend on the choice of the sound speed. The appropriate
sound speed to use here is the isothermal sound speed.

Figure~\ref{shockratios-isothermal} or its equivalent can then be used to
find the correction factor for the range of values of $u_0$ and $u_1$ which
seem reasonable.  Observational data on $u_0$ is scarce, but if it cannot be
further constrained, $10\pm 10$~km~s$^{-1}$ should cover most cases. For $u_1$
the question is which spectroscopic velocity to pick. Ultimately, for an
individual PN this question can only be answered through detailed
(photo-ionization or hydrodynamic) modelling of the PN, assuming it is
actually possible to produce a unique model for it. The approach of
\citet{1998A&A...329..265G}, who derive spatial velocity profiles using
photo-ionization models, may be useful here. Generic hydrodynamic modelling
may help establish what would be a good choice in general cases and I
understand that a project to do this is under way (Sch\"onberner, private
communication). Until that time the choice for $u_1$ will introduce
uncertainties in the method, as it has always done. It is important to
realize that the jump conditions are valid just before and after the front,
so the best choice would be a velocity as close as possible to the front.}

Round or elliptical PNe with attached shells are particularly suitable for
the expansion parallax method. Their inner bright rim is bounded by a shock,
and the presence of a surrounding ionized shell shows that this is not an
ionization front. For these almost round PNe, long slit spectroscopy can give
a good indication of the spatial and velocity structure, since it will be
quite similar in the directions perpendicular and parallel to the line of
sight.  This way a distinction can be made between the velocity of the rim
and the shell (the latter can be both higher and lower than that of the
rim). If possible it is best to choose species which lie closest to the outer
edge of the rim so as to avoid any velocity variations within the rim. For
these rims numerical models show that they are expanding into a region of low
velocity, typically lower than that of the original AGB wind. Choosing a
value for $u_0$ of around 5~km~s$^{-1}$ is a reasonable guess. In any case,
the corrections as derived in Sect.~2 only weakly depend on this choice.
Using the measured spectroscopic velocity, the ratio ${\cal R}$ can be
determined.

It would be interesting to derive an independent distance using the attached
shells, if available. If the long slit spectroscopy provides a velocity for
the shell, then the corrections from Sect.~4 can be used to find a
distance. Models show that the velocities of the shells are not as strongly
position dependent as those of the inner rims, something which is reflected
in the only mild ellipticities found observationally. This may make the outer
edges of shells actually better suited for distance determinations with the
expansion parallax method.

It is best to avoid areas and features well away from where a shock is
suspected. At least for a shock we can make a connection between the pattern
and material velocities. Away from these discontinuities, the two will also
differ, but there is no way to know how.

Bipolar PNe display a larger range of expansion velocities, which means that
it is harder to correct for inclination effects.  \citet{2002AJ....123.2666P}
state that the method is unusable for extreme bipolars, which seems too
pessimistic. However, a thorough understanding of the dynamics of the
complete PN is essential to be able to apply the method to these PNe, and in
most cases this information is not available. It would also be valuable to
have more radiation-hydrodynamic modelling of such systems.

\section{Conclusions}

Pattern velocities and the material (or bulk) velocities in a gas are not
necessarily the same. For discontinuities, such as shocks and ionization
fronts, the relation between the two can easily be derived using the jump
conditions across them. The pattern velocity is then found to be always
higher than the material velocity. Since measuring the expansion of PNe in
the sky is mostly done using sharp edges, which are associated with either
shocks or ionization fronts, a correction should be applied before
calculating the distance from the ratio of the two velocities. This
correction is typically larger for velocities of the order one or two times
the sound speed in the ionized material, which is actually what is measured
in most PNe.

Not using this correction will systematically underestimate the distances to
PNe. For the sample of PNe to which the expansion parallax method has been
applied successfully, the correction factors are around 1.2 to 1.3 for both
shocks and ionization fronts. Applying the corrections given in this paper
should lead to improved distance determinations to PNe.

\begin{acknowledgements} 
I like to thank Pierre Cox and Patrick Huggins for providing me with a value
for the equatorial expansion velocity of NGC 7027. My research has
been made possible by a fellowship of the Royal Netherlands Academy of Arts
and Sciences.
\end{acknowledgements} 

\bibliographystyle{aa} 
\bibliography{0485.bib} 

\end{document}